\documentclass[conference]{IEEEtran}
\ifCLASSINFOpdf
\else
\fi
\usepackage{amsmath}
\usepackage{amsfonts}

\usepackage{graphicx}
\usepackage{import}
\usepackage{color}
\usepackage{siunitx}

\usepackage{pgfplots} 
\usetikzlibrary{pgfplots.groupplots}
\usepgfplotslibrary{units}

\newcommand\copyrighttext{%
    \footnotesize \copyright~2018 IEEE. Personal use of this material is permitted. Permission from IEEE must be obtained for all other uses, in any current or future media, including reprinting/republishing this material for advertising or promotional purposes, creating new collective works, for resale or redistribution to servers or lists, or reuse of any copyrighted component of this work in other works. Link: \url{https://ieeexplore.ieee.org/abstract/document/7528065}}%
\newcommand\copyrightnotice{%
    \begin{tikzpicture}[remember picture,overlay]
    \node[anchor=south,yshift=10pt] at (current page.south) {\fbox{\parbox{\dimexpr\textwidth-\fboxsep-\fboxrule\relax}{\copyrighttext}}};
    \end{tikzpicture}%
}

\begin{document}
%
\title{The Adaptive Labeled Multi-Bernoulli Filter}

\author{\IEEEauthorblockN{Andreas Danzer, Stephan Reuter, Klaus Dietmayer}
\IEEEauthorblockA{Institute of Measurement, Control, and Microtechnology, Ulm University\\
Ulm, Germany\\
Email: \{andreas.danzer, stephan.reuter, klaus.dietmayer\}@uni-ulm.de}
}


%


\maketitle
\copyrightnotice

\begin{abstract}
	This paper proposes a new multi-Bernoulli filter called the Adaptive Labeled Multi-Bernoulli filter.
It combines the relative strengths of the known $\delta$-Generalized Labeled Multi-Bernoulli and the Labeled Multi-Bernoulli filter.
The proposed filter provides a more precise target tracking in critical situations, where the Labeled Multi-Bernoulli filter looses information through the approximation error in the update step.
In noncritical situations it inherits the advantage of the Labeled Multi-Bernoulli filter to reduce the computational complexity by using the LMB approximation.
\end{abstract}


%
\IEEEpeerreviewmaketitle

\section{Introduction}
The aim of multi-object tracking is the estimation of the number of objects as well as their
individual states based on noisy measurements, where missed detections and false alarms lead
to ambiguities in the track-to-measurement association. Further, adequate models for the appearance
and disappearance of objects are required. Approaches to tackle multi-object tracking
are Joint Probabilistic Data Association (JPDA) \cite{Bar-Shalom1988},
Multiple Hypotheses Tracking (MHT) \cite{Reid1979}, and the Random Finite Set (RFS) based 
multi-object Bayes filter \cite{Mahler2007}. 

Based on the mathematical tools of finite set statistics (FISST) \cite{Mahler2007}, 
several approximations of the multi-object Bayes filter have been proposed during the
last decade. The Probability Hypothesis Density (PHD) filter \cite{Mahler2003,Vo2006c,Sidenbladh2003,Vo2005}
and the Cardinalized PHD (CPHD) filter \cite{Mahler2007a,Vo2007} approximate the multi-object posterior
density by the first statistical moment and, in case of the CPHD filter, the cardinality distribution.
The Cardinality Balanced Multi-Target Multi-Bernoulli (CB-MeMBer) filter \cite{Vo2009}) approximates
the multi-object posterior using parameters of a multi-Bernoulli distribution and only propagates these
parameters. In \cite{Vo2013}, the class of labeled RFSs as well as the first analytic implementation
of the multi-object Bayes filter in form of the Generalized Labeled Multi-Bernoulli (GLMB) and the
$\delta$-GLMB filter are proposed. The $\delta$-GLMB filter is shown to outperform the approximations
of the multi-object Bayes filter and the incorporation of the track labels in the filtering step
significantly improves the track extraction in sequential Monte-Carlo (SMC) implementations \cite{Vo2013}.
The labeled Multi-Bernoulli (LMB) filter \cite{Reuter2014} efficiently approximates the $\delta$-GLMB filter by approximating
the posterior after each update step using an LMB distribution. In \cite{Reuter2014b}, the LMB filter
is shown to outperform PHD, CPHD, and multi-Bernoulli filters and to achieve almost the same accuracy as
the $\delta$-GLMB filter. Further, the LMB filter is successfully used for the real-time environment perception system of
the autonomous car of Ulm University \cite{Kunz2015,AutonomHP2014,AutonomVideo2014} based on radar, lidar, and
video sensors.

In \cite{Reuter2014}, the approximation error of the LMB filter with respect to the cardinality distribution
is illustrated in detail. Due to the assumption of statistically independent objects, the LMB representation
does not facilitate multi-modal cardinality distributions. In contrast to the $\delta$-GLMB filter which uses
multiple hypotheses to represent the data association uncertainty, the LMB filter represents the uncertainty
within the spatial distribution of each track. Hence, depending on merging and pruning thresholds applied for
the spatial distributions of the LMB filter, this representation may also lead to a loss of information.
Consequently, an adaptive multi-object tracking algorithm, which represents the tracks in $\delta$-GLMB representation in challenging
scenarios (e.g. objects are close by or track-to-measurement association is ambiguous) and uses the computationally
efficient LMB representation in all other scenarios, is expected to outperform the LMB filter and to adjust
the computational complexity to the complexity of the scenario.

In this contribution, the Adaptive Labeled Multi-Bernoulli (ALMB) filter is proposed which automatically switches
between an LMB and $\delta$-GLMB representation based on Kullback Leibler divergence \cite{Kullback1951} and
entropy \cite{Shannon1948}. The ALMB filter represents the multi-object posterior at each time step using
a set of LMB and $\delta$-GLMB distributions. Thus, only a subset of tracks is required to be represented in
$\delta$-GLMB form. The ALMB filter is compared to the $\delta$-GLMB and the LMB filter using two simulations.


This paper is organized as follows: First, the basics of (labeled) random finite
sets are outlined and the LMB filter as well as the $\delta$-GLMB filter are introduced. Section \ref{sec:switch_criteria}
introduces the switching criteria and the scheme of the ALMB filter is detailed in Section \ref{sec:almb}.
Finally, simulation results are shown in Section \ref{sec:results}.

\section{Background}
This sections summarizes multi-object tracking using random finite sets (RFS) and introduces 
the labeled multi-Bernoulli and the generalized labeled multi-Bernoulli RFS.

\subsection{Random Finite Sets}

An RFS is a finite-set-valued random variable with a random number of points, which are also random and unordered.
The RFS 
$		\text{X} = \{x^{(1)},\dots,x^{(N)}\} \subset \mathbb{X}$
represents the multi-object state $\text{X}$ with finite single-target state vectors $x^{(i)} \in \mathbb{X}$, where $\mathbb{X}$ is the state space.
Further, the RFS
$		\text{Z} = \{z^{(1)},\dots,z^{(M)}\} \subset \mathbb{Z}$
represents the multi-object observations with a random measurement $z^{(i)}$ out of the measurement space $\mathbb{Z}$.
The finite set statistics introduced in \cite{Mahler2007} are a powerful mathematical tool for dealing with RFSs.


\subsection{Multi-Bernoulli RFS}

A Bernoulli RFS is empty with a probability $1-r$ and has probability $r$ of being a singleton with a distribution $p$ defined on $\mathbb{X}$.
Its probability density is given by (see \cite{Mahler2007})
\begin{equation}
	\pi (\text{X}) = 
	\begin{cases}
		1 - r &\text{X} = \emptyset, \\
		r \cdot p(x) &\text{X} = \{x\}.
	\end{cases}
\end{equation}
The cardinality distribution is a Bernoulli distribution with parameter $r$.

A multi-Bernoulli RFS is the union of $M$ independent Bernoulli RFSs $\text{X}^{(i)}$, thus, $\text{X} = \bigcup_{i=1}^{M} \text{X}^{(i)}$ and is completely described by the parameter set $\{(r^{(i)},p^{(i)})\}_{i=1}^{M}$, where $r^{(i)}$ is the existence probability and $p^{(i)}$ the spatial distribution.

\subsection{Labeled Multi-Bernoulli RFS}

In a multi-object scenario, it is often required to estimate the identity of an object in addition to its current state.
For that reason, the class of labeled RFSs \cite{Vo2013} appends a label $\ell \in \mathbb{L}$ to each state state vector $x \in \mathbb{X}$. Thus, a labeled RFS is an RFS on $\mathbb{X} \times \mathbb{L}$ with state space $\mathbb{X}$ and finite label space $\mathbb{L}$.
In the following, labeled state vectors $\textbf{x}=(x,\ell)$ as well
as labeled RFSs $\textbf{X}$ are represented by bold letters.

Using the projection $\mathcal{L} : \mathbb{X} \times \mathbb{L} \rightarrow \mathbb{L}$ defined by ${\mathcal{L}((x,\ell)) = \ell}$,
the distinct label indicator 
\begin{equation}
	\Delta(\textbf{X}) = \delta_{\vert \textbf{X} \vert}(\vert \mathcal{L}(\textbf{X}) \vert),
\end{equation}
where $\mathcal{L}(\textbf{X}) = \{\mathcal{L}(\textbf{x}) : \textbf{x} \in \textbf{X}\}$ is the set of labels,
ensures that labels $\ell$ of a realization are distinct.

Similar to the multi-Bernoulli RFS, a labeled multi-Bernoulli (LMB) RFS is completely defined by the parameter set
$\{(r^{(\ell)},p^{(\ell)})\}_{\ell \in \mathbb{L}}$ and its density is given by (see \cite{Reuter2014})
\begin{equation}
	\boldsymbol \pi(\textbf{X}) = \Delta(\textbf{X}) w(\mathcal{L}(\textbf{X})) p^{\textbf{X}},
\end{equation}
where 
\begin{align}
	w(L) &= \prod\limits_{i \in \mathbb{L}}\left(1-r^{(i)}\right) \prod\limits_{\ell \in L}\dfrac{1_{\mathbb{L}}(\ell)r^{(\ell)}}{1-r^{(\ell)}}, \label{eq:lmb_weight}\\
	p(x,\ell) &= p^{(\ell)}(x).
\end{align}
The cardinality distribution of an LMB RFS is identical to the one of its unlabeled version and is given by
\begin{align}
\rho_{\text{LMB}}(n) &= \prod\limits_{i \in \mathbb{L}}\left(1-r^{(i)}\right) \sum\limits_{I \in \mathcal{F}_n(\mathbb{L})}
\prod\limits_{\ell \in I}\dfrac{1_{\mathbb{L}}(\ell)r^{(\ell)}}{1-r^{(\ell)}},
\label{eq:lmb_cardinality}
\end{align}
where $\mathcal{F}_n(\mathbb{L})$ denotes all subsets of $\mathbb{L}$ with exactly $n$ elements.

\subsection{$\delta$-Generalized Labeled Multi-Bernoulli RFS}

A generalized labeled multi-Bernoulli (GLMB) RFS \cite{Vo2013} is a labeled RFS with state space $\mathbb{X}$ and (discrete) label space $\mathbb{L}$ distributed according to
\begin{equation}
	\boldsymbol \pi(\textbf{X}) = \Delta(\textbf{X}) \sum\limits_{c \in \mathbb{C}} w^{(c)}(\mathcal{L}(\textbf{X})) \left[p^{(c)}\right]^{\textbf{X}},
\end{equation}
where $\mathbb{C}$ is a discrete index set and
\begin{align}
	\sum\limits_{L \subseteq \mathbb{L}} \sum\limits_{c \in \mathbb{C}} w^{(c)}(L) &= 1, \\
	\int\limits p^{(c)}(x,\ell)\mathrm{d}x &= 1.
\end{align}

A $\delta$-generalized labeled multi-Bernoulli ($\delta$-GLMB) RFS \cite{Vo2013} with state space $\mathbb{X}$ and (discrete) label space $\mathbb{L}$ is a special case of a generalized labeled multi-Bernoulli RFS with
\begin{align}
	\mathbb{C} &= \mathcal{F}(\mathbb{L}) \times \Xi, \\
	w^{(c)}(L) &= w^{(I,\xi)}\delta_I(L), \\
	p^{(c)} &= p^{(I,\xi)} = p^{(\xi)}
\end{align}
where $\mathcal{F}_n(\mathbb{L})$ denotes all subsets of $\mathbb{L}$, the discrete space $\Xi$ represents the history of track to measurement associations with realizations $\xi \in \Xi$ and $I$ is a set of track labels.
The density of a $\delta$-GLMB RFS is given by
\begin{equation}
	\boldsymbol \pi(\textbf{X}) = \Delta(\textbf{X}) \sum\limits_{(I,\xi) \in \mathcal{F}(\mathbb{L}) \times \Xi} w^{(I,\xi)}\delta_I(\mathcal{L}(\textbf{X})) \left[p^{(\xi)}\right]^{\textbf{X}}
	\label{dglmb_density}
\end{equation}
and its cardinality distribution follows
\begin{equation}
	\boldsymbol \rho_{\delta\text{-GLMB}}(n) = \sum\limits_{(I,\xi) \in \mathcal{F}_n(\mathbb{L}) \times \Xi} w^{(I,\xi)}.
	\label{eq:dglmb_cardinality}
\end{equation}

Obviously, an LMB RFS is a special case of a $\delta$-GLMB RFS with only one single component, i.e.
$p^{(\xi)}(x,\ell) = p^{(\ell)}(x)$:
\begin{equation}
	\boldsymbol \pi (\textbf{X}) = \Delta (\textbf{X}) \sum\limits_{I \in \mathcal{F}(\mathbb{L})} w^{(I)} \delta_I (\mathcal{L}(\textbf{X})) p^{\textbf{X}},
	\label{eq:dglmb_transformation}
\end{equation}
where the weights of the components follow \eqref{eq:lmb_weight}.

\section{$\delta$-Generalized Labeled Multi-Bernoulli Filter}

The $\delta$-GLMB filter was introduced in \cite{Vo2013}, where it is shown that GLMBs and $\delta$-GLMBs are conjugate priors \footnote{Note: the number of components increases due to the association uncertainty.} with respect to the prediction and update equations of the multi-object Bayes filter \cite{Mahler2007}.

\subsection{Prediction}

The prediction of a $\delta$-generalized labeled multi-Bernoulli of the form \eqref{dglmb_density} to the time of the next measurement
is given by
\begin{align}
	\boldsymbol \pi_+(\textbf{X}) = \Delta(\textbf{X}) \!\!\! \sum\limits_{(I_+,\xi) \in \mathcal{F}(\mathbb{L}_+) \times \Xi} \!\!\!w_+^{(I_+,\xi)} \delta_{I_+}(\mathcal{L}(\textbf{X})) \left[ p_+^{(\xi)} \right]^{\textbf{X}},
	\label{eq:dglmb_prediction}
\end{align}
where 
\begin{align}
	w_+^{(I_+,\xi)} &= w_B(I_+ \cap \mathbb{B}) w_S^{(\xi)}(I_+ \cap \mathbb{L}) ,\\
	p_+^{(\xi)}(x,\ell) &= 1_{\mathbb{L}}(\ell)p_S^{(\xi)}(x,\ell) + 1_{\mathbb{B}}(\ell)p_B(x,\ell) ,\\
	p_S^{(\xi)}(x,\ell) &= \dfrac{\langle p_S(\cdot,\ell) f(x \vert \cdot,\ell),p^{(\xi)}(\cdot,\ell) \rangle}{\eta_S^{(\xi)}(\ell)}, \\
	\eta_S^{(\xi)}(\ell) &= \int \langle p_S(\cdot,\ell) f(x \vert \cdot,\ell),p^{(\xi)}(\cdot,\ell) \rangle \mathrm{d}x, \\
	w_S^{(\xi)}(L) &= \left[\eta_S^{(\xi)}\right]^L \sum\limits_{I \subseteq \mathbb{L}} 1_I(L) \left[q_S^{(\xi)}\right]^{I-L} w^{(I,\xi)},\\
	q_S^{(\xi)} &= \left\langle q_S(\cdot,\ell),p^{(\xi)}(\cdot,\ell) \right\rangle.
	\label{eq:dglmb_prediction_end}
\end{align}
In (\ref{eq:dglmb_prediction})-(\ref{eq:dglmb_prediction_end}), $w_B(\cdot)$ is the weight of the birth labels $I_+ \cap \mathbb{B}$ and $w_S^{(\xi)}(\cdot)$ of the surviving labels $I_+ \cap \mathbb{L}$.
Further, $p_B(\cdot,\cdot)$ is the density of new-born objects and $p_S^{(\xi)}(\cdot,\cdot)$ of surviving objects, depending on the transition density $f(x \vert \cdot,\ell)$ weighted by the probability of survival $p_S(\cdot,\ell)$ and the prior density $p^{(\xi)}(\cdot,\ell)$.
Besides, $\langle f , g \rangle = \int f(x)g(x)dx$ denotes the inner product, $\eta_S^{(\xi)}(\ell)$ is a normalization constant and $q_S(\cdot,\ell) = 1 - p_S(\cdot,\ell)$ the probability that a track disappears.

\subsection{Update}
The posterior density after the measurement update of \eqref{eq:dglmb_prediction} is again a $\delta$-GLMB RFS given by
\begin{align}
	\boldsymbol \pi(\textbf{X} \vert \text{Z}) = \Delta(&\textbf{X}) \sum\limits_{(I_+,\xi) \in \mathcal{F}(\mathbb{L}_+) \times \Xi} \sum\limits_{\theta \in \Theta} w^{(I_+,\xi,\theta)}(\text{Z}) \nonumber\\
		&\times \delta_{I_+}(\mathcal{L}(\textbf{X})) \left[ p^{(\xi,\theta)}(\cdot \vert \text{Z}) \right]^{\textbf{X}}
	\label{eq:dglmb_update}
\end{align}
where 
\begin{align}
	w&^{(I_+,\xi,\theta)}(\text{Z}) \propto \delta_{\theta^{-1}(\{0:\vert \text{Z} \vert\})}(I_+) w_+^{(I_+,\xi)} \left[\eta_{\text{Z}}^{(\xi,\theta)}\right]^{I_+},\\
	p&^{(\xi,\theta)}(x,\ell \vert \text{Z}) = \dfrac{p_+^{(\xi)}(x,\ell) \psi_{\text{Z}}(x,\ell;\theta)}{\eta_{\text{Z}}^{(\xi,\theta)}(\ell)}, \\
	\eta&_{\text{Z}}^{(\xi,\theta)}(\ell) = \left\langle p_+^{(\xi)}(\cdot,\ell),\psi_{\text{Z}}(\cdot,\ell;\theta) \right\rangle, \\
	\psi&_{\text{Z}}(x,\ell;\theta) = \delta_0(\theta(\ell)) q_D(x,\ell) \nonumber \\
	&\hspace{1.8cm} +(1-\delta_0(\theta(\ell))) \dfrac{p_D(x,\ell) g(z_{\theta(\ell)} \vert x,\ell)}{\kappa(z_{\theta(\ell)})}.
	\label{eq:dglmb_update_end}
\end{align}
In (\ref{eq:dglmb_update})-(\ref{eq:dglmb_update_end}), $\theta \in \Theta : I_+ \rightarrow \{0,1,\dots,\vert Z \vert\}$ associates track labels to measurements, where $\theta(i) = 0$ represents a missing detection and $\theta(i) = \theta(j) > 0$ implies $i \equiv j$.
Note, the posterior sets of track labels correspond to the predicted sets of track labels, i.e. $I = I_+$.
Here, $w^{(I_+,\xi,\theta)}$ is the updated weight of a hypothesis $(I_+,\xi,\theta)$.
Further, $\eta_{\text{Z}}^{(\xi,\theta)}(\ell)$ is a normalization constant and $\psi_{\text{Z}}(x,\ell;\theta)$ is the measurement likelihood.
The likelihood depends on the probability of a missing detection $q_D(x,\ell) = 1 - p_D(x,\ell)$ at $(x,\ell)$ and the spatial likelihood $g(z_{\theta(\ell)} \vert x,\ell)$ weighted by the detection probability $p_D(x,\ell)$ at $(x,\ell)$.
$\kappa(z_{\theta(\ell)}) = \lambda_c c(z)$ models the intensity of Poisson clutter.

\section{Labeled Multi-Bernoulli Filter}

The Labeled Multi-Bernoulli (LMB) filter was proposed in \cite{Reuter2014} and is intended as a
fast and accurate approximation of the $\delta$-GLMB filter. While an LMB RFS is a conjugate prior
with respect to the prediction equations, the filter update requires a transformation to $\delta$-GLMB form
and a subsequent approximation.

\subsection{Prediction}
For a multi-object posterior LMB RFS with parameter set ${\boldsymbol \pi = \{(r^{(\ell)},p^{(\ell)})\}_{\ell \in \mathbb{L}}}$ on $\mathbb{X} \times \mathbb{L}$ and a multi-object LMB birth density $\boldsymbol \pi_B = \{(r_B^{(\ell)},p_B^{(\ell)})\}_{\ell \in \mathbb{B}}$ on $\mathbb{X} \times \mathbb{B}$, the multi-object prediction is also an LMB RFS with state space $\mathbb{X}$ and finite label space $\mathbb{L}_+ = \mathbb{B} \cup \mathbb{L}$ and is given by 
\begin{equation}
	\boldsymbol \pi_+\left(\textbf{X}\right) = \left\{\left( r_{+,S}^{(\ell)},p_{+,S}^{(\ell)} \right)\right\}_{\ell \in \mathbb{L}} \cup \left\{\left( r_{B}^{(\ell)},p_{B}^{(\ell)} \right)\right\}_{\ell \in \mathbb{B}},
	\label{eq:lmb_prediction}
\end{equation}
where
\begin{align}
	r_{+,S}^{(\ell)} & = \eta_S(\ell) r^{(\ell)},\\
	p_{+,S}^{(\ell)} & = \dfrac{\langle p_S(\cdot,\ell) f(x \vert \cdot,\ell), p(\cdot,\ell) \rangle}{\eta_S(\ell)},\\
	\eta_S(\ell) &= \int \langle p_S(\cdot,\ell) f(x \vert \cdot,\ell),p(\cdot,\ell) \rangle \mathrm{d}x.
	\label{eq:lmb_prediction_end}
\end{align}
In (\ref{eq:lmb_prediction})-(\ref{eq:lmb_prediction_end}), $p_S(\cdot,\ell)$ denotes the state dependent survival probability and $f(x \vert \cdot,\ell)$ the single target transition density for track $\ell$.
Further, $\eta_S(\ell)$ is a normalization constant.

\subsection{Update}
Since an LMB RFS is not a conjugate prior with respect to the measurement update of the multi-object Bayes filter, the LMB
filter update transforms the predicted LMB RFS to a corresponding $\delta$-GLMB RFS using \eqref{eq:dglmb_transformation}
and subsequently applies the $\delta$-GLMB update. In order to reduce the computational complexity, the
LMB components and the measurements are partitioned into approximately statistically independent groups
(see \cite{Reuter2014} for a detailed explanation).

In order to close the LMB filter recursion, an approximation of the updated $\delta$-GLMB distribution using an
LMB RFS is required, i.e.
\begin{equation}
	\boldsymbol \pi \left(\textbf{X}\right) \approx \widetilde{\boldsymbol \pi} \left(\textbf{X}\right) = \left\{\left( r^{(\ell)},p^{(\ell)} \right)\right\}_{\ell \in \mathbb{L}}.
	\label{eq:lmb_approximation}
\end{equation}
The parameters of the LMB RFS are obtained from the updated $\delta$-GLMB components using
\begin{align}
	r&^{(\ell)} = \sum\limits_{(I_+,\theta) \in \mathcal{F}(\mathbb{L}_+) \times \Theta_{I_+}} w^{(I_+,\theta)}(Z) 1_{I_+}(\ell), \label{eq:lmb_r} \\
	p&^{(\ell)}(x) = \dfrac{1}{r^{(\ell)}} \!\! \sum\limits_{(I_+,\theta) \in \mathcal{F}(\mathbb{L}_+) \times \Theta_{I_+}}\!\! w^{(I_+,\theta)}(Z) 1_{I_+}(\ell) p^{(\theta)}(x,\ell|\rm{Z}), \label{eq:lmb_p} 
\end{align}
where $\Theta_{I_+}$ denotes the space of mappings. The weights $w^{(I_+,\theta)}$ and the densities $p^{(\theta)}$
are computed using (\ref{eq:dglmb_update})-(\ref{eq:dglmb_update_end}) by setting $\xi=\emptyset$.

The LMB approximation does not affect the spatial distributions of the individual tracks but due to the
assumption of statistically independent tracks within an LMB RFS, the cardinality distribution of the
approximation may differ whereas the mean cardinality is identical (see \cite{Reuter2014} for additional details).

\section{Switching Criteria}
\label{sec:switch_criteria}
The aim of the ALMB filter is to switch automatically between the LMB approximation, 
facilitating a fast propagation of the density, and the more accurate $\delta$-GLMB density.
In the following, two switching criteria are introduced where the Kullback-Leibler criterion
evaluates the approximation error of the LMB approximation and the Entropy criterion
considers the data association uncertainty.

\subsection{Kullback-Leibler Criterion}
\label{sec:kl_criterion}

In \cite{Reuter2014}, it was shown that the LMB approximation loses information about the cardinality distribution.
Thus, an intuitive way to detect the information loss, is to examine the difference between the posterior
cardinality distribution of the $\delta$-GLMB RFS and its LMB approximation.
The Kullback-Leibler (KL) divergence \cite{Kullback1951} is a measure to compare two (discrete) probability distributions $P$ and $Q$:
\begin{equation}
	D_{KL}(P \Vert Q) = \sum\limits_{i} P(i) \cdot \log\dfrac{P(i)}{Q(i)},
	\label{eq:kullback_leibler}
\end{equation}
where $Q$ is the approximation of $P$ and the $i$-th term is zero, if $P(i) = 0$ since
\begin{equation}
	\lim\limits_{x \to 0} x\log(x) = 0.
\end{equation}

The cardinality distributions of the $\delta$-GLMB RFS $\boldsymbol \rho_{\delta\text{-GLMB}}(n)$ and its LMB approximation
$\boldsymbol \rho_{\text{LMB}}(n)$ are given by (\ref{eq:dglmb_cardinality}) and (\ref{eq:lmb_cardinality}),
hence the Kullback-Leibler criterion calculates
\begin{equation}
	D_{KL}(\boldsymbol \pi (\textbf{X})) = D_{KL}(\boldsymbol \rho_{\delta\text{-GLMB}} \Vert \boldsymbol \rho_{\text{LMB}}),
\end{equation}
where $\boldsymbol \pi (\textbf{X})$ denotes the updated $\delta$-GLMB RFS which facilitates the calculation of
its LMB.
For $D_{KL} = 0$, the cardinality distributions are identical, i.e. the LMB approximation causes no information loss in the cardinality distribution.
$D_{KL} > 0$ implies that $\boldsymbol \rho_{\text{LMB}}$ differs from $\boldsymbol \rho_{\delta\text{-GLMB}}$, where a large value of $D_{KL}$ can be interpreted as big difference between the cardinality distributions, i.e., a large approximation error.


\subsection{Entropy Criterion}
\label{sec:entropy_criterion}

The $\delta$-GLMB RFS comprises several hypotheses to capture the data association uncertainties,
whereas the LMB RFS captures the association uncertainty within the spatial distributions of a individual track.
Depending on the parameters and the post-processing of the spatial distributions, the $\delta$-GLMB representation
is in general more accurate in challenging situations. For example, two Gaussian components obtained
by different track-to-measurement associations may be merged in a Gaussian Mixture (GM) LMB filter which results in
a loss of information compared to the $\delta$-GLMB representation holding the two associations in different
hypotheses. 

The Kullback-Leibler criterion does not detect challenging situations with ambiguous data association if the cardinality distributions are identical.
Hence, a measure for the data association uncertainty is required which enables a switching to
the more accurate $\delta$-GLMB representation in these situations.
The entropy \cite{Shannon1948} is a measure of unpredictability of information content and is widely
used in information theory. In this contribution, the entropy is used to evaluate the track-to-measurement
association. Following \cite{Shannon1948}, the entropy is
\begin{equation}
	H(P) = -\sum\limits_{i} P(x_i) \log P(x_i),
\end{equation}
where $P(x_i)$ is the probability that the event $x_i$ occurs.
Obviously, for small or large values $P(x_i)$, the entropy is small.
This fact can be used to detect ambiguous data associations.

The assocation matrix of tracks to measurements is given by
\begin{equation}
	A = 
	\begin{pmatrix}
		r^{(\ell_1,z_1)} & r^{(\ell_1,z_2)} & \cdots & r^{(\ell_1,z_m)} \\
		r^{(\ell_2,z_1)} & r^{(\ell_2,z_2)} & \cdots & r^{(\ell_2,z_m)} \\
		\vdots & \vdots & \ddots & \vdots \\
		r^{(\ell_n,z_1)} & r^{(\ell_n,z_2)} & \cdots & r^{(\ell_n,z_m)}
	\end{pmatrix},
\end{equation}
where
\begin{equation}
	r^{(\ell_i,z_j)} = \sum\limits_{(I_+,\theta) \in \mathcal{F}(\mathbb{L_+}) \times \Theta_{I_+}} w^{(I_+,\theta)}(Z) 1_{I_+}(\ell_i) \delta_{\theta(\ell_i)}(j)
\end{equation}
is the probability that track $\ell_i$ is assigned to measurement $z_j$.
Further, $\theta \in \Theta_{I_+} : I_+ \rightarrow \{0,1,\ldots,\vert Z \vert\}$ is a mapping of labels to measurements in such a way that $\theta(i) = \theta(j) > 0$ implies $i \equiv j$.

An unambiguous assignment of measurement $z_j$ is characterized by the column vector $a_j$ with one value $r^{(\ell_i,z_j)} \approx 1$ and all other values $r^{(\ell_k,z_j)} \approx 0, k = 1,2,\ldots,i-1,i+1,\ldots,n$.
With the above-mentioned property of entropy, such a column vector results in a small value $H(a_j)$.
Hence, the entropy for a distribution $\boldsymbol\pi (\textbf{X})$ is
\begin{equation}
	H(\boldsymbol\pi (\textbf{X})) = \sum\limits_{j=1}^m H(a_{j}),
\end{equation}
where a small value indicates an unambiguous data association and a large value represents an uncertain track-to-measurement assignment.


\section{The Adaptive Labeled Multi-Bernoulli Filter}
\label{sec:almb}
\label{sec:almb_components}
The $\delta$-GLMB filter of \cite{Vo2013} is shown in \cite{Reuter2014b,Reuter2015b,Beard2016} to outperform the
LMB filter \cite{Reuter2014} in challenging scenarios, e.g. containing closely spaced objects in combination
with missed detections and false alarms, at the cost of a significantly higher computational complexity.
The main idea of the Adaptive Labeled Multi-Bernoulli (ALMB) filter proposed in this section is to combine
the advantages of the $\delta$-GLMB filter in critical situations with the efficiency of the LMB filter.
Thus, the ALMB filter uses the $\delta$-GLMB distribution to represent tracks in critical situations as
accurate as possible and uses LMB distributions for the representation of all other tracks. 

Using the assumption, that well separated objects are statistically independent of each other, the ALMB filter
uses several independent multi-object distributions in LMB and $\delta$-GLMB form to represent the environment, 
i.e.
\begin{align}
	\boldsymbol \pi_{\delta\text{-GLMB}}(\textbf{X}^{(\delta)}) &= \left\{ \boldsymbol \pi_{\delta\text{-GLMB}}^{(i)}(\textbf{X}^{(i)})\right\}_{i=1}^{n_{\delta}},\\
	\boldsymbol \pi_{\text{LMB}}(\textbf{X}^{(L)}) &= \left\{ \boldsymbol \pi_{\text{LMB}}^{(i)}(\textbf{X}^{(i)})\right\}_{i=1}^{n_L},
\end{align}
where $n_{\delta}$ is the number of $\delta$-GLMB distributions and $n_L$ denotes the number of LMB distributions.
Consequently, the multi-object posterior density is given by the set
\begin{align}
	\boldsymbol \pi(\textbf{X}) = \left\{ \boldsymbol \pi_{\delta\text{-GLMB}}(\textbf{X}^{(\delta)}), \boldsymbol \pi_{\text{LMB}}(\textbf{X}^{(L)})\right\}.\label{eq:almb_density}
\end{align}

\begin{figure}[!t]
	\centering
	\def\svgwidth{\columnwidth}
	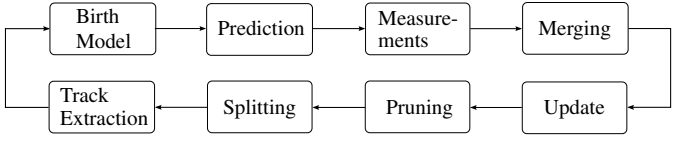
	\caption{Scheme of the proposed ALMB filter.}
	\label{fig:almb_procedure}
\end{figure}

Fig. \ref{fig:almb_procedure} illustrates the scheme of the ALMB filter which propagates the
density \eqref{eq:almb_density} over time. Obviously, each component of the ALMB filter 
is required to be able to handle both representations, LMB and $\delta$-GLMB. In the following,
the individual components of the ALMB filter are presented in detail.

\subsection{Birth Model}
The birth model is responsible for initializing new tracks. The ALMB filter may use a static birth model
\cite{Vo2013} requiring known birth locations or an adaptive birth model \cite{Reuter2014,Reuter2014b}
which facilitates the appearance of objects anywhere in the state space. Due to the structure of the
ALMB filter, several new-born objects may not be represented using a single LMB distribution since this
would require an additional splitting of distributions before the measurement update. Hence, each new-born
object $\ell_B$ is represented by an individual LMB distribution consisting of a single component
\begin{align}
\boldsymbol \pi_{B}^{(\ell_B)} = \left\{r_{B}^{(\ell_B)}(z),p_{B}^{(\ell_B)}(x) \right\}.
\end{align}

\subsection{Prediction}
In the prediction step, the ALMB filter predicts each distribution $\boldsymbol \pi_{\delta\text{-GLMB}}^{(i)}(\textbf{X}^{(i)})$, $i=1,\ldots,n_{\delta}$, using the standard $\delta$-GLMB prediction equations. Further, the standard LMB prediction is applied for 
each $\boldsymbol \pi_{\text{LMB}}^{(i)}(\textbf{X}^{(i)})$, $i=1,\ldots,n_L$.

\subsection{Measurements}
Since the ALMB filter holds multiple multi-object distributions at the same time, not every received measurement
affects each multi-object density. Hence, the measurements module performs a gating procedure for the distributions
and the set of measurements. Obviously, this gating procedure is important for a parallel execution of the
update step in the manner of \cite{Reuter2014}.

The measurements module always performs the assignment of observations $z^{(i)} \in \text{Z}$ based on the
LMB distribution. Consequently, a $\delta$-GLMB RFS $ \boldsymbol \pi_{\delta\text{-GLMB}}^{(i)}(\textbf{X}^{(i)})$ 
has to be approximated by an LMB RFS $\widetilde{\boldsymbol \pi}_{\text{LMB}}^{(i)}(\textbf{X}^{(i)})$ according to (\ref{eq:lmb_approximation}). 
This temporary approximation facilitates a faster observation to distribution association. Since the
spatial distribution of the LMB approximation is equivalent \cite{Reuter2014}, the temporary approximation
does not influence the gating result.

After that, the gating examines if a received observation $z^{(i)}$ affects the track $\ell$ of the LMB distribution with
\begin{equation}
	d_{MHD}(\hat{z}_+^{(\ell)},z^{(i)}) < \sqrt{\gamma_z}, 
\end{equation}
whereby $\hat{z}_+^{(\ell)}$ is the predicted measurement of track $\ell$ and $\gamma_z$ is the gating distance threshold.
The value of $\gamma_z$ depends on the desired $\sigma$-gate of the confidence interval and can be calculated using the inverse Chi-squared cumulative distribution.

\subsection{Merging}
\label{sec:merging}
Due to the new-born objects represented by individual LMB RFSs and the prediction of the existing multi-object densities,
it is possible that distributions influence each other in the update step. Hence the merging combines all multi-object
densities with common measurements.

The measurements module assigned a set of measurements $\text{Z}^{(i)}$ to the predicted multi-object density $\boldsymbol \pi_+^{(i)}$.
Consequently, two predicted densities $\boldsymbol \pi_+^{(i)}$ and $\boldsymbol \pi_+^{(j)}$ with common measurements
\begin{equation}
	\text{Z}^{(i)} \cap \text{Z}^{(j)} \neq \emptyset
\end{equation}
have to be merged into a single multi-object distribution $\boldsymbol \pi_+^{(i,j)}$.
The merging itself depends on the representations of $\boldsymbol \pi_+^{(i)}$ and $\boldsymbol \pi_+^{(j)}$
and is introduced in the following.

\subsubsection{Merging of LMBs}
\label{sec:merging_lmb}
The merging of two LMB RFSs is used as in \cite{Reuter2014} after the parallel group update step.
Two LMB densities $\boldsymbol \pi^{(i)}(\textbf{X}^{(i)})$ and $\boldsymbol \pi^{(j)}(\textbf{X}^{(j)})$ are merged to the distribution
\begin{equation}
	\boldsymbol \pi^{(i,j)}(\textbf{X}^{(i,j)}) = \boldsymbol \pi^{(i)}(\textbf{X}^{(i)}) \cup \boldsymbol \pi^{(j)}(\textbf{X}^{(j)}).
	\label{eq:merging_lmb}
\end{equation}

\subsubsection{Merging of $\delta$-GLMBs}
\label{sec:merging_dglmb}

The merging of two $\delta$-GLMB RFS is not as simple as the LMB merging, because all combinations of the hypotheses of the two RFSs have to be considered.
To calculate the combined $\delta$-GLMB RFS, each component of $\boldsymbol \pi_{\delta\text{-GLMB}}^{(i)}(\textbf{X}^{(i)})$ has to be multiplied with each component of $\boldsymbol \pi_{\delta\text{-GLMB}}^{(j)}(\textbf{X}^{(j)})$ resulting in
\begin{align}
	\boldsymbol \pi^{(i,j)}(&\textbf{X}^{(i,j)}) = \Delta(\textbf{X}^{(i,j)}) \nonumber\\
&	\times \sum\limits_{(I,\xi) \in \mathcal{F}(\mathbb{L}^{(i)}) \times \Xi^{(i)}} w^{(I,\xi)} \delta_{I}(\mathcal{L}(\textbf{X}^{(i)}) \left[p^{(\xi)}\right]^{\textbf{X}^{(i)}} \\
&	\times \sum\limits_{(\tilde{I},\tilde{\xi}) \in \mathcal{F}(\mathbb{L}^{(j)}) \times \Xi^{(j)}} w^{(\tilde{I},\tilde{\xi})} 
\delta_{\tilde{I}}(\mathcal{L}(\textbf{X}^{(j)})) \left[p^{(\tilde{\xi})}\right]^{\textbf{X}^{(j)}}.
		\label{eq:merging_dglmb}
\end{align}
Hence, the merged number of components is given by the product of the individual number of components.

\subsubsection{Merging of an LMB with a $\delta$-GLMB}
\label{sec:merging_lmb_dglmb}

The merging of an LMB RFS with a $\delta$-GLMB RFS always results in a \mbox{$\delta$-GLMB}.
First, the LMB $ \boldsymbol \pi_{\text{LMB}}^{(i)}(\textbf{X})$ is transformed into a corresponding $\delta$-GLMB
$ \boldsymbol \pi_{\delta\text{-GLMB}}^{(i)}(\textbf{X})$ using (\ref{eq:dglmb_transformation}).
Due to the fact that both densities are now in $\delta$-GLMB form, the $\delta$-GLMB merging
according to (\ref{eq:merging_dglmb}) is used to merge the densities.

\subsection{Update}

According to the representation of the predicted multi-object density, the ALMB filter performs either a $\delta$-GLMB or a LMB update,
which are given by (\ref{eq:dglmb_update}) and (\ref{eq:lmb_approximation}), respectively. Observe: in case of the LMB update
it is required to store the resulting $\delta$-GLMB density of the update step in addition for the following steps.

After the update, the ALMB filter uses the criteria presented in Section \ref{sec:switch_criteria} to decide
whether a switching of the multi-object representation is necessary or not.
An updated LMB RFS indicates a noncritical situation before the update, but it is possible that the correction step
changed this fact. Therefore, the filter examines the cardinality distributions of the LMB approximation $\widetilde{\boldsymbol \pi} (\textbf{X})$ and the $\delta$-GLMB posterior ${\boldsymbol \pi} (\textbf{X})$ using the Kullback-Leibler criterion (see Section \ref{sec:kl_criterion}) to detect an information loss and Entropy criterion (see \ref{sec:entropy_criterion}) to handle ambiguous
track-to-measurement associations. If one of the criteria detects a critical situation, i.e. the KL divergence or the entropy exceed an application-specific threshold, the filter replaces the LMB approximation
by the $\delta$-GLMB RFS obtained during the filter update and propagates the incorporated tracks using the \mbox{$\delta$-GLMB} filter
in the next filter cycle.

If a loss of information caused a switching, the filter uses the KL divergence to examine whether the critical situation is solved.
Otherwise, the Entropy criterion is used.
This implies that only the criterion which detected the critical situation, can trigger switching back to propagating
a set of tracks using an LMB RFS.
Once the KL divergence or the entropy fall below the threshold, a challenging situation is resolved.

\subsection{Pruning}

Both, the LMB update and the $\delta$-GLMB update, produce components with negligible influence.
To reduce the computational cost, the pruning removes these components.

The LMB pruning removes all tracks $\ell$ with marginal existence probability $r^{(\ell)}$, so the resulting LMB RFS is
\begin{equation}
	\widetilde{\boldsymbol \pi}_{\text{LMB}}^{(i)} = \left\{\left( r^{(\ell)},p^{(\ell)} \right): r^{(\ell)} > \mu_r \right\}_{\ell \in \mathbb{L}^{(i)}},
\end{equation}
where $\mu_r$ represents the application dependent minimum existence probability.

The $\delta$-GLMB pruning removes all hypotheses $(I,\xi,\theta)$ with insignificant weight $w^{(I,\xi,\theta)}$, which leads to
\begin{equation}
	\widetilde{\boldsymbol \pi}_{\delta\text{-GLMB}}^{(i)}  = \left\{(I,\xi,\theta) : w^{(I,\xi,\theta)} > \mu_w \right\}_{I \in \mathcal{F}(\mathbb{L}^{(i)})}
\end{equation}
using the threshold $\mu_w$.

\subsection{Splitting}

In the course of time, it is possible that tracks in a density move apart, so that the RFS can be splitted into multiple smaller distributions of the same type.
The splitting uses the grouping algorithm of \cite{Reuter2014}.
Therefore, a $\delta$-GLMB RFS temporary is approximated by an LMB RFS according to (\ref{eq:lmb_approximation}) during the splitting procedure. Then, the partitioning scheme is applied to find a possible splitting of the RFS.

The module splits an LMB distribution $\boldsymbol \pi^{(i)}$ in multiple new densities $\boldsymbol \pi^{(j)}$ such that
\begin{equation}
	\boldsymbol \pi^{(i)} = \bigcup\limits_{j=1}^{N} \boldsymbol \pi^{(j)},
\end{equation}
where $N$ is the number of identified object groups.

The $\delta$-GLMB splitting uses the labels of tracks in the groups to create several new $\delta$-GLMB RFS.
The splitted $\delta$-GLMB densities $\boldsymbol \pi^{(j)}$ only approximate the original distribution
$\boldsymbol \pi^{(i)}$, because during the splitting, hypotheses containing labels of two different groups,
are divided in new hypotheses, which only contain the labels according to the new distribution.
Since the influence of tracks from different groups is marginal, the occurred approximation error is negligible.

\subsection{Track Extraction}

The track extraction decides whether a track with label $\ell$ exists or not by using the existence probability $r^{(\ell)}$.
An LMB density
\begin{equation}
	\boldsymbol \pi^{(j)} = \left\{\left( r^{(\ell)},p^{(\ell)} \right) \right\}_{\ell \in \mathbb{L}^{(j)}}
\end{equation}
implicitly contains the existence probability.
According to \cite{Reuter2014}, the extraction of the track is
\begin{equation}
	\hat{\textbf{X}} = \left\{ (\hat{x},\ell) : r^{(\ell)} > \vartheta \right\},
	\label{eq:4_almb_track_extraction}
\end{equation}
where the parameter $\vartheta$ is an application specific threshold and ${\hat{x} = \arg\limits_{x} \max p^{(\ell)}(x)}$.

Since a $\delta$-GLMB density does not contain the existence probability, the module has to calculate this value.
Following \cite{Reuter2014}, the existence probability for a track $\ell$ is given by
\begin{equation}
	r^{(\ell)} = \sum\limits_{(I,\theta) \in \mathcal{F}(\mathbb{L}) \times \Theta_I} w^{(I,\theta)}(\text{Z}) 1_{I}(\ell),
\end{equation}
where $w^{(I,\theta)}$ is the weight of the corresponding hypothesis $(I,\theta)$.
Afterwards, the extraction uses (\ref{eq:4_almb_track_extraction}) to choose existing tracks.

\section{Results}
\label{sec:results}
This section evaluates the ALMB filter and compares it with the LMB filter \cite{Reuter2014} and the $\delta$-GLMB filter \cite{Vo2013}.
For the evaluation, a gaussian mixture (GM) implementation is used.

The scenario consists of two targets on a two dimensional region $[-1000,1000]\text{m} \times [-1000,1000]\text{m}$.
The target state $x_k = [p_{x,k}, \dot{p}_{x,k}, p_{y,k}, \dot{p}_{y,k}]^T$ comprises the position and velocity in $x$ and $y$ direction.
Measurements are noisy vectors $z_k = [z_{x,k}, z_{y,k}]^T$. The clutter measurements are uniformly distributed over the measurement
space and their number follows a Poisson distribution with mean value $\lambda_c$.

The state model is a standard constant velocity model where the standard deviation of the process noise for the velocity in $x$ and $y$ direction is given by $\sigma_v^2 = 5 \text{m}/\text{s}^2$. The cycle time of the sensor is $T=1~\rm{s}$ and the standard deviation
of the sensor measurements consisting of $x$ and $y$ positions is $\sigma_{\varepsilon} = 10 \text{m}$.

The survival probability of the targets is state independent and given by $p_{S,k} = 0.99$, the detection probability is ${p_{D,k} = 0.98}$.
Furthermore, the birth densities are two multi-Bernoulli RFS $\boldsymbol \pi_{B}^{(i)} = \{r_{B}^{(i)}(z),p_{B}^{(i)}(x \vert z)\}_{i = 1}^2$, where ${r_B^{(1)} = r_B^{(2)} = 0.05}$, $p_B^{(i)} = \mathcal{N}(x;m_B^{(i)},P_B)$ with $m_B^{(1)} = [-1000,0,0,0]^T$, ${m_B^{(2)} = [1000,0,0,0]^T}$ and ${P_B = \text{diag}([10,10,10,10]^T)^2}$.

The thresholds for an automatic switching between an LMB and $\delta$-GLMB representation are $10^{-4}$ for the Kullback-Leibler criterion and $0.5$ for the Entropy criterion. 
In a $\delta$-GLMB density, the number of components is limited to $50$ and the pruning removes all components with a weight below $10^{-5}$.
In an LMB density, all tracks with an existence probability below $0.01$ are pruned.
For both densities, the threshold in the track extraction is set to $0.5$ and an extracted track is represented by the gaussian component with the highest weight.

\begin{figure}[!t]
	\centering
	\resizebox{\columnwidth}{!}{\input{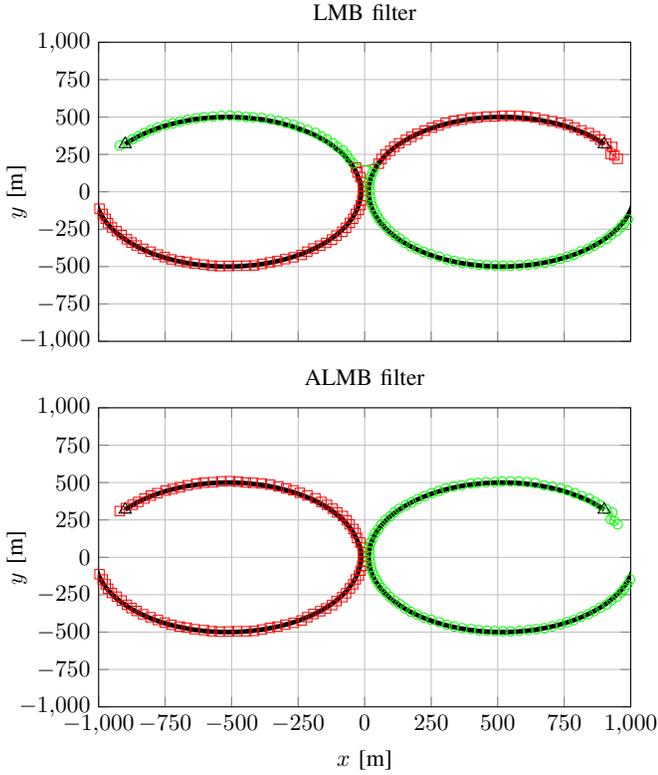}}
	\caption{Ground truth trajectories of the two objects (black lines with endposition marked by a triangle), the estimated trajectories (circles/squares) of the LMB filter (above) and the result of the ALMB filter (below).}
	\label{fig:results_filters}
\end{figure}

Figure \ref{fig:results_filters} shows the true trajectories together with the tracking result of a single run.
Obviously, the LMB filter can not handle the situation in the region $[-100,100]\text{m} \times [0,250]\text{m}$.
In this critical situation, the data association of tracks to measurements is uncertain and the LMB filter erroneously switches the track labels. In contrast, the ALMB filter successfully detects the critical situation using the criteria from Section \ref{sec:switch_criteria} and uses the $\delta$-GLMB representation until the ambiguity is resolved. As a result, the ALMB filter can deal with such situations and does not switch the track labels.

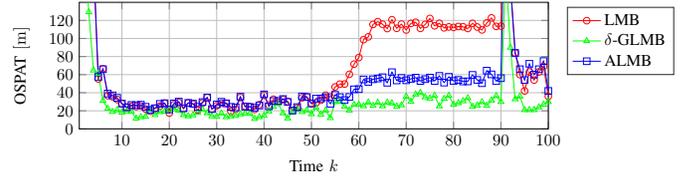
\begin{figure}[!t]
	\centering
	\resizebox{\columnwidth}{!}{
%
%
\definecolor{mycolor1}{rgb}{1,0,0}%
\definecolor{mycolor2}{rgb}{0,1,0}%
\definecolor{mycolor3}{rgb}{0,0,1}%
\begin{tikzpicture}

\begin{axis}[%
width=10cm,
height=2.7cm,
at={(2.526in,1.284in)},
scale only axis,
xmin=1,
xmax=100,
xtick={0,10,...,100},
xlabel={Time $k$},
xmajorgrids,
ymin=0,
ymax=140,
ytick={0,20,...,120},
y unit=\si{\meter},
ylabel={OSPAT},
ymajorgrids,
axis background/.style={fill=white},
legend style={at={(0,1)},xshift=0.1cm,yshift=-0.1cm,legend cell align=left,align=left,draw=white!15!black},
legend pos=outer north east,
]
\addplot [color=mycolor1,solid,mark=o,mark options={solid}]
  table[row sep=crcr]{%
1	300\\
2	300\\
3	298.695763560971\\
4	186.127780273222\\
5	55.2869784026349\\
6	66.4728012136165\\
7	36.7615779362442\\
8	35.6732399818208\\
9	32.5015102010415\\
10	27.9591192540133\\
11	23.8753016110739\\
12	24.9928736818188\\
13	24.7392226905312\\
14	25.2901763593396\\
15	22.0093358707744\\
16	19.4913895766049\\
17	22.8180942723932\\
18	25.1467145185871\\
19	28.3958391902911\\
20	17.5121427793283\\
21	26.7122734076678\\
22	29.1831901515108\\
23	22.2963698570622\\
24	28.7379765114089\\
25	27.4123429704004\\
26	20.9789380817924\\
27	28.6630779072464\\
28	34.6634525224665\\
29	21.9141931966384\\
30	25.6256952382537\\
31	28.5244814133759\\
32	28.2467070141072\\
33	19.4123457860601\\
34	21.8496691342225\\
35	34.662122187302\\
36	25.0459917337631\\
37	24.6832577840609\\
38	21.5313274838014\\
39	26.8007270085145\\
40	37.2094168241489\\
41	26.3920644126065\\
42	32.7931923771381\\
43	29.4963418094261\\
44	32.5386370857183\\
45	30.2744429630999\\
46	20.2683977358715\\
47	23.1297875665149\\
48	34.3796582527947\\
49	35.6742420694025\\
50	28.3705070698356\\
51	28.7100487149489\\
52	31.7687867566044\\
53	37.2748769787561\\
54	35.2938988483705\\
55	46.3484340475212\\
56	49.5143336530708\\
57	49.598762350685\\
58	60.3572474553299\\
59	71.6103487465919\\
60	78.8491148344119\\
61	98.7090562381718\\
62	101.765657849179\\
63	115.529263754936\\
64	118.603039927733\\
65	115.784057209846\\
66	110.987009160406\\
67	120.745117535957\\
68	111.251944024512\\
69	115.906438470152\\
70	109.412512025975\\
71	116.693615381125\\
72	117.951314238781\\
73	111.184800684363\\
74	115.520469375756\\
75	122.138835563653\\
76	114.006668443256\\
77	116.756922644587\\
78	111.777701764685\\
79	112.778962046266\\
80	112.235047265652\\
81	113.282909455377\\
82	112.944440944392\\
83	111.064462931277\\
84	116.711222508239\\
85	112.673023584646\\
86	109.495731422813\\
87	117.127101792391\\
88	122.743397770026\\
89	113.381040626781\\
90	113.725735390831\\
91	300\\
92	162\\
93	84\\
94	60\\
95	42\\
96	63\\
97	54\\
98	63\\
99	69\\
100	36\\
};
\addlegendentry{LMB};

\addplot [color=mycolor2,solid,mark=triangle,mark options={solid}]
  table[row sep=crcr]{%
1	287.005171392885\\
2	208.32498368927\\
3	129.22489051877\\
4	64.738500515695\\
5	61.0540113319914\\
6	31.2008431900821\\
7	22.440184891011\\
8	19.0357353290921\\
9	20.7585400800474\\
10	18.09783967752\\
11	19.5077577065171\\
12	19.1518391128827\\
13	11.5726540420842\\
14	13.0684618840197\\
15	13.6890911223354\\
16	21.4795722486236\\
17	15.4921448939078\\
18	18.731684347494\\
19	19.1255940767259\\
20	20.9043535409489\\
21	21.8638753932619\\
22	20.3706570966681\\
23	15.4195128009993\\
24	14.0731143237028\\
25	15.179985642555\\
26	18.027886366779\\
27	21.2864814275944\\
28	17.5299221342061\\
29	14.0389138857762\\
30	13.5935787867304\\
31	17.2930591680086\\
32	13.0974463800059\\
33	14.0383409523316\\
34	14.5320154411335\\
35	16.2626142140331\\
36	17.1232965315187\\
37	16.8248517743801\\
38	11.2241787059549\\
39	13.11660127814\\
40	14.7217034184958\\
41	19.5384371537465\\
42	24.4250376170384\\
43	27.9932772053284\\
44	17.414581149367\\
45	11.7724455564563\\
46	18.7474949226683\\
47	21.8181664250582\\
48	18.8840296388105\\
49	20.5539613562775\\
50	16.4579820698313\\
51	23.5305153168686\\
52	15.888092531732\\
53	20.4333040269578\\
54	12.1464901642501\\
55	29.8590372063862\\
56	30.3141859900621\\
57	24.8429815945491\\
58	20.5060896889294\\
59	26.765036306159\\
60	27.0225846418208\\
61	25.2613443465662\\
62	29.3515797863096\\
63	25.6822907573584\\
64	29.1229181163623\\
65	25.2172491901202\\
66	29.2225137959162\\
67	28.3451681044755\\
68	24.8891469986024\\
69	26.2600791884639\\
70	35.8041586957209\\
71	30.1403702539806\\
72	38.208781749853\\
73	39.9672701765085\\
74	35.0345497621394\\
75	33.7003050014654\\
76	36.7159890289104\\
77	25.1281776482669\\
78	33.3177295496816\\
79	36.7654502105991\\
80	26.8976621038666\\
81	28.6271138359194\\
82	30.7475673335886\\
83	34.1789486600974\\
84	24.5972856317388\\
85	26.0515724193528\\
86	29.7025123021461\\
87	25.6927090584848\\
88	31.5689208364658\\
89	36.4855647748047\\
90	29.9600630017724\\
91	174\\
92	90\\
93	33\\
94	36\\
95	21\\
96	21\\
97	21\\
98	24\\
99	27\\
100	30\\
};
\addlegendentry{$\delta$-GLMB};

\addplot [color=mycolor3,solid,mark=square,mark options={solid}]
  table[row sep=crcr]{%
1	300\\
2	300\\
3	298.695763560971\\
4	186.619475361998\\
5	57.6074360314001\\
6	66.0082552634015\\
7	39.0167416529028\\
8	37.1200204860272\\
9	34.4110314484178\\
10	27.9585550382182\\
11	23.8754416887573\\
12	26.4560693417373\\
13	26.2130260569783\\
14	26.7496686216939\\
15	24.4415833037739\\
16	20.9524518564286\\
17	22.8124844392608\\
18	27.5843325438885\\
19	28.3956411938187\\
20	21.4128258396846\\
21	28.1823978424556\\
22	30.1593273940659\\
23	22.7880568671116\\
24	28.7381066717556\\
25	27.9035089363245\\
26	24.4139423565189\\
27	29.6464641823075\\
28	34.6695088999546\\
29	21.9141843763386\\
30	27.1033505729611\\
31	29.9974676807391\\
32	28.2467016863068\\
33	24.3078211324923\\
34	23.798890376558\\
35	35.6347151554081\\
36	24.5481892878789\\
37	24.200142091985\\
38	23.4887579188293\\
39	25.819617639888\\
40	38.181852039822\\
41	24.9336822309092\\
42	30.8461448314639\\
43	30.9640775274087\\
44	33.031469275396\\
45	29.7879817670353\\
46	20.2683942840391\\
47	24.1034313972337\\
48	35.8479099447779\\
49	35.1819110353591\\
50	27.9561336536418\\
51	33.5229852055591\\
52	33.6196473465263\\
53	35.0331788426904\\
54	27.794637696592\\
55	37.6770780234159\\
56	35.1359294162558\\
57	31.8894537971325\\
58	35.3600948438562\\
59	43.7518046165718\\
60	43.9747084586582\\
61	54.6800316696113\\
62	51.9802800185036\\
63	54.8586560344948\\
64	55.736532588606\\
65	57.1527774392257\\
66	51.0393141081383\\
67	62.737625526685\\
68	53.583436755479\\
69	56.3731614031094\\
70	53.743984213434\\
71	55.3219300080323\\
72	56.6793612030771\\
73	53.0089783948151\\
74	55.0125948871612\\
75	56.745930695943\\
76	52.975891382362\\
77	58.6043225274598\\
78	51.1817451236186\\
79	57.5599735732704\\
80	53.0510966899417\\
81	54.175691988605\\
82	52.458059173517\\
83	52.5428591825432\\
84	59.6947057725509\\
85	54.1398697520261\\
86	50.5456280977995\\
87	64.4112275782482\\
88	60.2733358813801\\
89	52.8678990240866\\
90	56.1230898782988\\
91	300\\
92	159\\
93	84\\
94	66\\
95	54\\
96	72\\
97	60\\
98	66\\
99	75\\
100	42\\
};
\addlegendentry{ALMB};

\end{axis}
\end{tikzpicture}%
}
	\caption{OSPAT distances of order $p = 1$ and cut-off $c = 300$ for GM implementation with $\lambda_c = 50$ and $p_D = 0.98$ (averaged over $100$ MC runs).}
	\label{fig:results_ospat}
\end{figure}

The OSPAT distances \cite{Ristic2011} in Fig. \ref{fig:results_ospat} illustrate the difference between the LMB, $\delta$-GLMB and ALMB filters.
In noncritical situations (time $k < 55$), the performance of the LMB and ALMB filter is identical since the ALMB filter uses the LMB representation. However, LMB and ALMB perform slightly worse than the $\delta$-GLMB filter which is expected due to the approximations
in the update step.
In challenging situations with data association uncertaintities (time $k > 50 \wedge k < 65$), the ALMB and $\delta$-GLMB filter outperform the LMB filter.
Obviously, the ALMB filter can handle the situation due to the propagation of multiple hypotheses in most of the Monte Carlo runs.
In contrast, the LMB filter loses too much information due to the LMB approximation and almost always switches the track labels.
At time $k = 91$, the OSPAT increases for all filters due to the disappearance of both tracks and the short delay
until both tracks are abandoned. Figure \ref{fig:results_filters} obviously illustrates this fact.
After a certain time, the estimation of the tracks matches to the ground truth resulting in a declining OSPAT distance.

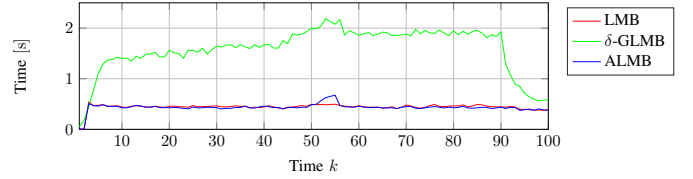
\begin{figure}[!t]
	\centering
	\resizebox{\columnwidth}{!}{
%
%
\definecolor{mycolor1}{rgb}{1,0,0}%
\definecolor{mycolor2}{rgb}{0,1,0}%
\definecolor{mycolor3}{rgb}{0,0,1}%
\begin{tikzpicture}

\begin{axis}[%
width=10cm,
height=2.7cm,
at={(2.526in,1.284in)},
scale only axis,
xmin=1,
xmax=100,
xtick={0,10,...,100},
xlabel={Time $k$},
xmajorgrids,
ymin=0,
ymax=2.5,
y unit=\si{\second},
ylabel={Time},
ymajorgrids,
axis background/.style={fill=white},
legend style={at={(0,1)},xshift=0.1cm,yshift=-0.1cm,legend cell align=left,align=left,draw=white!15!black},
legend pos=outer north east,
]
\addplot [color=mycolor1,solid]
  table[row sep=crcr]{%
1	0.00935444\\
2	0.0014836\\
3	0.53887237\\
4	0.46771944\\
5	0.47545357\\
6	0.48530976\\
7	0.45615259\\
8	0.45903996\\
9	0.46589124\\
10	0.44195106\\
11	0.43232201\\
12	0.45698164\\
13	0.46611576\\
14	0.46010895\\
15	0.43602942\\
16	0.4399981\\
17	0.45639941\\
18	0.44279972\\
19	0.43431064\\
20	0.4554768\\
21	0.45952175\\
22	0.44827123\\
23	0.4451616\\
24	0.43573319\\
25	0.47065578\\
26	0.4465569\\
27	0.44865833\\
28	0.45201667\\
29	0.4579206\\
30	0.46469623\\
31	0.44573262\\
32	0.44322176\\
33	0.45152195\\
34	0.46461943\\
35	0.45425859\\
36	0.45795584\\
37	0.42583285\\
38	0.43789495\\
39	0.42715923\\
40	0.44383435\\
41	0.46645392\\
42	0.44462307\\
43	0.4225817\\
44	0.40784651\\
45	0.43752573\\
46	0.44102725\\
47	0.45467965\\
48	0.47023079\\
49	0.45517571\\
50	0.49494375\\
51	0.48779996\\
52	0.49143507\\
53	0.48580833\\
54	0.49259504\\
55	0.49591403\\
56	0.47460067\\
57	0.45770404\\
58	0.44910584\\
59	0.44279541\\
60	0.46910596\\
61	0.44543509\\
62	0.43849037\\
63	0.43797521\\
64	0.43659794\\
65	0.42286962\\
66	0.42515141\\
67	0.44625621\\
68	0.41085902\\
69	0.43762836\\
70	0.47964048\\
71	0.46260632\\
72	0.43097176\\
73	0.42008365\\
74	0.43145821\\
75	0.47040975\\
76	0.49209433\\
77	0.45226502\\
78	0.45049945\\
79	0.45709859\\
80	0.46211605\\
81	0.46226936\\
82	0.48333315\\
83	0.44606174\\
84	0.46260979\\
85	0.49081874\\
86	0.48427948\\
87	0.45847529\\
88	0.44972701\\
89	0.44273649\\
90	0.4450632\\
91	0.43682661\\
92	0.45321426\\
93	0.44330476\\
94	0.39237603\\
95	0.39776961\\
96	0.3854867\\
97	0.38129118\\
98	0.38139014\\
99	0.37759539\\
100	0.38737834\\
};
\addlegendentry{LMB};

\addplot [color=mycolor2,solid]
  table[row sep=crcr]{%
1	0.05522275\\
2	0.1714377\\
3	0.45396221\\
4	0.77455298\\
5	1.10112535\\
6	1.28950316\\
7	1.37045892\\
8	1.38108238\\
9	1.42534347\\
10	1.40162215\\
11	1.40865081\\
12	1.34994316\\
13	1.44359962\\
14	1.40497342\\
15	1.48046952\\
16	1.48867332\\
17	1.51861616\\
18	1.42940383\\
19	1.43250967\\
20	1.56313047\\
21	1.5136071\\
22	1.4932066\\
23	1.40604895\\
24	1.56381155\\
25	1.55726477\\
26	1.52235122\\
27	1.60427079\\
28	1.48697874\\
29	1.57366199\\
30	1.64667784\\
31	1.62344151\\
32	1.66658334\\
33	1.67066017\\
34	1.61487178\\
35	1.67232143\\
36	1.63019874\\
37	1.59751448\\
38	1.6700379\\
39	1.61957956\\
40	1.66868453\\
41	1.6810605\\
42	1.60536797\\
43	1.60708081\\
44	1.75083925\\
45	1.69570317\\
46	1.88866862\\
47	1.86142371\\
48	1.88358698\\
49	1.93180026\\
50	1.99856638\\
51	1.98510594\\
52	2.03961592\\
53	2.18670527\\
54	2.12618357\\
55	2.07782805\\
56	2.16775565\\
57	1.86424467\\
58	1.95166359\\
59	1.90771277\\
60	1.8550851\\
61	1.88800356\\
62	1.9114429\\
63	1.85596941\\
64	1.88067884\\
65	1.92192424\\
66	1.90698571\\
67	1.85268303\\
68	1.85844504\\
69	1.86340757\\
70	1.88057689\\
71	1.84871466\\
72	1.97738357\\
73	1.88688387\\
74	1.95443896\\
75	1.90436222\\
76	1.89674586\\
77	1.96451967\\
78	1.90430503\\
79	1.853777\\
80	1.94648265\\
81	1.87104733\\
82	1.93585636\\
83	1.87973573\\
84	1.93535028\\
85	1.95701925\\
86	1.9024831\\
87	1.81013339\\
88	1.87149555\\
89	1.83733376\\
90	1.92495295\\
91	1.28741722\\
92	1.07706814\\
93	0.88887081\\
94	0.84691587\\
95	0.70705957\\
96	0.64211772\\
97	0.59196808\\
98	0.56320289\\
99	0.57473343\\
100	0.58622125\\
};
\addlegendentry{$\delta$-GLMB};

\addplot [color=mycolor3,solid]
  table[row sep=crcr]{%
1	0.00521519\\
2	0.00132863\\
3	0.50784838\\
4	0.46866063\\
5	0.45879199\\
6	0.487828\\
7	0.45527885\\
8	0.45477178\\
9	0.45491233\\
10	0.4251273\\
11	0.42631051\\
12	0.44694783\\
13	0.47478522\\
14	0.46046222\\
15	0.44266132\\
16	0.43655543\\
17	0.45467481\\
18	0.42796968\\
19	0.42816224\\
20	0.42663787\\
21	0.43379465\\
22	0.42606864\\
23	0.41440466\\
24	0.4061062\\
25	0.44044862\\
26	0.42287602\\
27	0.42711052\\
28	0.43564095\\
29	0.43438184\\
30	0.42679215\\
31	0.40355732\\
32	0.41682491\\
33	0.41488858\\
34	0.43297742\\
35	0.43251933\\
36	0.4588771\\
37	0.42483016\\
38	0.41915203\\
39	0.44461048\\
40	0.45801886\\
41	0.4486178\\
42	0.43064209\\
43	0.43380229\\
44	0.40314376\\
45	0.43064687\\
46	0.43092712\\
47	0.42658977\\
48	0.43718478\\
49	0.43450019\\
50	0.48632642\\
51	0.4953261\\
52	0.56606154\\
53	0.62562549\\
54	0.6518746\\
55	0.67403542\\
56	0.46988403\\
57	0.44790488\\
58	0.44173349\\
59	0.43736702\\
60	0.45794317\\
61	0.43847531\\
62	0.43159864\\
63	0.43221113\\
64	0.43926197\\
65	0.42061344\\
66	0.42369483\\
67	0.42672377\\
68	0.41053828\\
69	0.4355475\\
70	0.44886909\\
71	0.45027797\\
72	0.41733845\\
73	0.40973215\\
74	0.4287747\\
75	0.44124485\\
76	0.45078593\\
77	0.43886863\\
78	0.42245975\\
79	0.4396317\\
80	0.42453234\\
81	0.43309498\\
82	0.45571254\\
83	0.42276559\\
84	0.41874541\\
85	0.42598328\\
86	0.43423601\\
87	0.43603738\\
88	0.41094254\\
89	0.42327802\\
90	0.43824249\\
91	0.41336085\\
92	0.43386775\\
93	0.43253854\\
94	0.37501136\\
95	0.38779948\\
96	0.40034499\\
97	0.37671425\\
98	0.4037374\\
99	0.3817186\\
100	0.37685336\\
};
\addlegendentry{ALMB};

\end{axis}
\end{tikzpicture}%
}
	\caption{Computation time of the LMB, $\delta$-GLMB and ALMB filter (averaged over $100$ MC runs).}
	\label{fig:results_time}
\end{figure}

Figure \ref{fig:results_time} shows the computation time of the compared filters.
Obviously, the LMB and ALMB filter significantly outperform the $\delta$-GLMB.
The execution time of the ALMB filter is almost the same as of the LMB filter.
Only in critical situations ($k > 50 \wedge k < 65$), the ALMB filter needs more time for the calculation, but nevertheless, it outperforms the \mbox{$\delta$-GLMB} filter in such situations.
In \cite{Hoang2015}, a fast implementation of the $\delta$-GLMB filter is proposed.
This implemenation reduces the execution time of the $\delta$-GLMB filter, but would also speed up the ALMB filter in critical situations.
The ALMB filter represents tracks by partitioned multi-Bernoulli RFSs resulting from the merging and splitting module.
Since the groups are assumed to be independent, the prediction, measurements, update, pruning and track extraction modules can be performed in parallel, which also speeds up the algorithm.

\begin{figure}[!t]
	\centering
	\resizebox{\columnwidth}{!}{\input{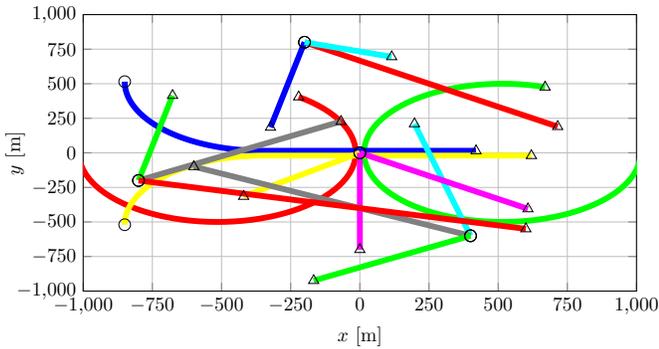}}
	\caption{Ground truth trajectories of a scenario with up to 16 objects, where the start position is marked by an circle and the end position by a triangle.}
	\label{fig:results_big_scen_ground_truth}
\end{figure}	

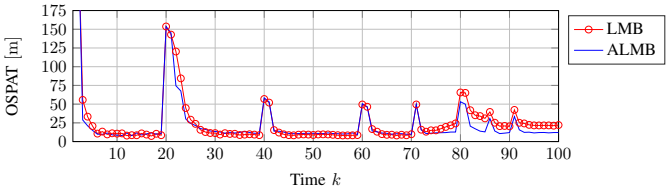
\begin{figure}[!t]
	\centering
	\resizebox{\columnwidth}{!}
	{
%
%
\definecolor{mycolor1}{rgb}{1,0,0}%
\definecolor{mycolor2}{rgb}{0,0,1}%
\begin{tikzpicture}

\begin{axis}[%
width=10cm,
height=2.7cm,
at={(2.526in,1.284in)},
scale only axis,
xmin=1,
xmax=100,
xtick={0,10,...,100},
xlabel={Time $k$},
xmajorgrids,
ymin=0,
ymax=175,
ytick={0,25,...,175},
y unit=\si{\meter},
ylabel={OSPAT},
ymajorgrids,
axis background/.style={fill=white},
legend style={at={(1,1)},xshift=-0.1cm,yshift=-0.1cm,legend cell align=left,align=left,draw=white!15!black},
legend pos=outer north east
]
\addplot [color=mycolor1,solid,mark=o,mark options={solid}]
  table[row sep=crcr]{%
1	300\\
2	300\\
3	55.6446330870695\\
4	33.2792464803552\\
5	20.7522440588999\\
6	10.6391947180886\\
7	13.3105805985332\\
8	9.58295024515636\\
9	11.0648145323581\\
10	10.5813062157157\\
11	10.9392112573466\\
12	8.00518518567757\\
13	8.51892324813167\\
14	8.43024830523548\\
15	10.9036024683761\\
16	9.34822201034836\\
17	7.46299861707653\\
18	10.4056031847926\\
19	8.51649622922016\\
20	153.699755458059\\
21	142.8341945587\\
22	120.276520644708\\
23	84.4165724993076\\
24	44.7811349884562\\
25	29.2270950928062\\
26	23.6286801799389\\
27	15.7291193389447\\
28	12.7440700967385\\
29	11.4440939008124\\
30	11.29211363073\\
31	9.35099619603226\\
32	11.1847918166005\\
33	10.2237418638116\\
34	10.4022655693482\\
35	8.77168810817201\\
36	9.27408479020588\\
37	9.80262592085333\\
38	9.29467130575525\\
39	8.58757444018311\\
40	56.8602025322851\\
41	52.0576286900911\\
42	15.3086848584639\\
43	12.011906136057\\
44	9.7043772561837\\
45	8.61515507369409\\
46	8.34129735522266\\
47	9.21889390597827\\
48	9.50508055800728\\
49	9.32618389401045\\
50	8.83096420442165\\
51	9.48440635861308\\
52	9.62674631929598\\
53	9.50408259750959\\
54	9.02266300701641\\
55	8.51417499997349\\
56	8.17832903030319\\
57	8.27868093304254\\
58	8.37719264478814\\
59	8.66002370393905\\
60	49.5162752244396\\
61	46.4990037028873\\
62	16.6871628471927\\
63	13.2003972972946\\
64	9.94381933656892\\
65	9.03611172534912\\
66	9.18041420070447\\
67	8.16065467579407\\
68	9.10876228125079\\
69	8.40856722971024\\
70	9.79874560962249\\
71	49.5390747920347\\
72	15.8922751104042\\
73	13.1740478206339\\
74	15.0161482279628\\
75	14.9897755559469\\
76	17.1256863568021\\
77	19.5650284288103\\
78	21.6588160812271\\
79	24.2200802151991\\
80	65.5687783110484\\
81	65.0252200431736\\
82	41.861317410975\\
83	35.1607003122115\\
84	34.0248475276164\\
85	31.0063193356957\\
86	39.5686383227844\\
87	25.1790552996179\\
88	20.5990213720595\\
89	20.8761012238247\\
90	20.3756204012123\\
91	42.2862977680006\\
92	25.10128852128\\
93	24.2111983305722\\
94	22.3250341210325\\
95	21.6221881663085\\
96	21.5557826611717\\
97	21.5934224687549\\
98	21.6154966941544\\
99	21.2498496965361\\
100	22.0815722331635\\
};
\addlegendentry{LMB};

\addplot [color=mycolor2,solid,mark=none,mark options={solid}]
  table[row sep=crcr]{%
1	300\\
2	300\\
3	28.8874468842252\\
4	21.0550832081712\\
5	14.9844155778296\\
6	9.87540817753066\\
7	9.62920830563204\\
8	11.3910805942693\\
9	8.97330229746837\\
10	8.57720277991717\\
11	8.05303745801385\\
12	11.4820278726256\\
13	7.88874411765141\\
14	11.5950852246001\\
15	8.86236080403485\\
16	10.2970996085317\\
17	11.5608605655065\\
18	9.87883394408517\\
19	10.9743278775017\\
20	153.465529587265\\
21	145.130145248846\\
22	74.5340215045876\\
23	67.5712220057429\\
24	30.2836371771672\\
25	24.2187238524883\\
26	21.4098289582712\\
27	17.2728323082487\\
28	16.3448388093608\\
29	14.1239220530642\\
30	13.4617247224844\\
31	12.8220574173592\\
32	11.0289957530911\\
33	12.4450089989225\\
34	11.4235371186463\\
35	12.2555616485521\\
36	13.2721646354733\\
37	12.1217989464694\\
38	11.090493520397\\
39	11.1069696324445\\
40	58.3808488081416\\
41	53.0577385762966\\
42	13.8733826313317\\
43	13.8768042100929\\
44	12.6382556471923\\
45	10.9005001997838\\
46	10.5078057012868\\
47	10.7262658317709\\
48	10.9021171822852\\
49	11.0481057882428\\
50	10.9968406248948\\
51	10.8901960647614\\
52	10.6227817875945\\
53	11.532671715062\\
54	10.7563136367764\\
55	10.5006735617053\\
56	10.123228404866\\
57	10.2852673687746\\
58	11.5539158297411\\
59	9.41618600369565\\
60	50.5145876311543\\
61	46.1393347220721\\
62	14.1735690676491\\
63	12.9337015675946\\
64	11.4027943422933\\
65	11.057811203638\\
66	9.90172284704245\\
67	10.4626957803462\\
68	10.5521209581493\\
69	9.78872935704964\\
70	9.70419413176034\\
71	49.418116593835\\
72	15.6177049722356\\
73	10.6238463887494\\
74	11.1736468283835\\
75	11.6760412254851\\
76	11.8230723776022\\
77	12.004226200905\\
78	12.7053409602987\\
79	12.7294430646107\\
80	53.1649311727005\\
81	49.9399727805028\\
82	20.5930073431329\\
83	17.0974880753391\\
84	13.9204877781582\\
85	12.9280485981014\\
86	30.7029766142876\\
87	14.1071501874643\\
88	10.6527241774942\\
89	11.2732301078718\\
90	12.0067733109412\\
91	33.3459024318574\\
92	15.683510421337\\
93	12.2826872378615\\
94	12.5649150902626\\
95	11.5585530893378\\
96	12.0427826019382\\
97	12.1592232027727\\
98	11.628879347318\\
99	12.056495626627\\
100	12.0654297038597\\
};
\addlegendentry{ALMB};

\end{axis}
\end{tikzpicture}%
}
	\caption{OSPAT distances of order $p = 1$ and cut-off $c = 300$ for GM implementation with $\lambda_c = 25$ and $p_D = 0.98$ (averaged over $100$ MC runs).}
	\label{fig:results_big_scen_ospat}
\end{figure}

In a second example, the performance of the ALMB filter is evaluated in a scenario with many targets.
Figure \ref{fig:results_big_scen_ground_truth} illustrates the scenario with up to 16 objects involving birth and death of objects and considering missed detections and clutter measurements. The OSPAT distances in Figure \ref{fig:results_big_scen_ospat} show that the ALMB filter always performs same or better than the LMB filter.

\section{Conclusion}
This paper has proposed a new efficient multi-target tracking filter based on a Labeled Multi-Bernoulli and $\delta$-Generalized Multi-Bernoulli filter.
The proposed Adaptive Labeled Multi-Bernoulli filter combines a low computational complexity of the LMB filter with the accuracy of the $\delta$-GLMB filter.
With the Kullback-Leibler distance and an intuitive interpretation of the entropy, the filter uses simple mathematical tools to detect challenging situations in a tracking scenario.
Since the criteria do not depend on the representation of the spatial distributions, the principles of the ALMB filter
may also be used in sequential Monte Carlo implementations as well as the recently published Gamma Gaussian Inverse Wishart
implementation \cite{Beard2016}.
The modular structure of the ALMB filter further facilitates the replacement of individual components and the extension of the filter with new features.


\section*{Acknowledgment}
This work is supported by the German Research Foundation
(DFG) within the Transregional Collaborative Research Center SFB/TRR 62
"Companion-Technology for Cognitive Technical Systems".



\bibliographystyle{IEEEtran}
\bibliography{IEEEabrv,mrm}
%
%
%

\end{document}